\documentclass[3p,times, onecolumn]{elsarticle}
\usepackage{physics}
\usepackage{siunitx}
\usepackage[dvipsnames]{xcolor}
\usepackage{csquotes}
\usepackage{amssymb}
\usepackage{amsmath}
\usepackage{pifont}
\usepackage{acro}
\DeclareAcronym{EMS}{
    short = EMS,
    long  = Energy Management System,
}
\DeclareAcronym{FLC}{
    short = FLC,
    long  = Fuzzy Logic Controller,
}
\DeclareAcronym{COP}{
    short = COP,
    long  = Coefficient of Performance,
}
\DeclareAcronym{SOC}{
    short = SOC,
    long  = State of Charge,
}
\DeclareAcronym{PV}{
    short = PV,
    long  = Photovoltaics,
}
\DeclareAcronym{RES}{
    short = RES,
    long  = Renewable Energy Source,
}
\DeclareAcronym{BESS}{
    short = BESS,
    long  = Battery Energy Storage System,
}
\DeclareAcronym{MPC}{
    short = MPC,
    long  = Model Predictive Control,
}
\DeclareAcronym{RBC}{
    short = RBC,
    long  = Rule-Based Control,
}
\DeclareAcronym{RL}{
    short = RL,
    long  = Reinforcement Learning,
}
\DeclareAcronym{LOH}{
    short = LOH,
    long  = Level of Hydrogen,
}
\DeclareAcronym{DER}{
    short = DER,
    long  = Distributed Energy Resources,
}
\DeclareAcronym{LHV}{
    short = LHV,
    long  = Lower Heating Value,
}
\DeclareAcronym{CEMS}{
    short = CEMS,
    long  = Comfort Energy Management System,
}
\DeclareAcronym{PEMFC}{
    short = PEMFC,
    long  = Proton Exchange Membrane Fuel Cell,
}
\DeclareAcronym{ECDF}{
    short = ECDF,
    long  = Empirical Cumulative Distribution Function,
}

\graphicspath{{figures/}}
\usepackage{lipsum}
\usepackage{booktabs}
\usepackage{float}
\usepackage{comment}
\usepackage{subcaption}
\usepackage{graphicx}
\usepackage{tikz}
\usepackage{framed} 
\usepackage{multicol} 
\usepackage{nomencl} 
\usepackage{enumitem}
\usepackage[framemethod=TikZ]{mdframed}
\usepackage{hyperref}
\hypersetup{hidelinks}
\usepackage{tabularx}
\usepackage{eurosym}
\DeclareSIUnit{\EUR}{\text{\euro}}
\usepackage{amssymb}
\usepackage{todonotes}
\usepackage{algorithm}
\usepackage[noend]{algpseudocode}
\algdef{SE}[SUBALG]{Indent}{EndIndent}{}{\algorithmicend\ }%
\algtext*{Indent}
\algtext*{EndIndent}

\usepackage{makecell}
\usepackage{natbib} 
\usepackage{multirow}
\usepackage{array}
\usepackage{diagbox}
\definecolor{KITgreen}{RGB}{0,150,130}
\newcolumntype{C}[1]{>{\centering\arraybackslash}p{#1}}
\newcommand{\good}{\textcolor{PineGreen}{\raisebox{-0.5ex}{\Large $\mathbf{+}$}}} \newcommand{\avg}{\textcolor{Dandelion}{\raisebox{-0.5ex}{\Large $\bullet$}}} \newcommand{\poor}{\textcolor{Maroon}{\raisebox{-0.5ex}{\Large \textbf{--}}}}
\usepackage{tablefootnote}

\journal{Energy and Buildings}

\begin{document}

\begin{frontmatter}

\newpage
\title{Comfort energy management system for electrified buildings using hydrogen for seasonal storage}

\author{Jovana Vu\v{c}ini\'c\corref{corauthor}}
\ead{jovana.vucinic@kit.edu}
\author{Felix Langner}
\author{Erfan Tajalli-Ardekani}
\author{Marvin Dorn}
\author{Simon Waczowicz}
\author{Ralf Mikut}
\author{J\"org Matthes}
\author{H\"useyin~K.~\c{C}akmak}
\author{Veit Hagenmeyer}
\address{Karlsruhe Institute of Technology, Institute for Automation and Applied Informatics, Eggenstein-Leopoldshafen, Germany}

\cortext[corauthor]{Corresponding author}

\begin{abstract}
\acresetall

Flexible loads and integrated energy storage in electrified buildings can align volatile renewable energy use with consumption.
While battery energy storage systems provide short-term flexibility, hydrogen storage enables seasonal energy shifting.
However, since hydrogen systems are expensive, coupling a fuel cell with a heat pump can reduce the required size of the hydrogen system and, therefore, investment costs.
To fully exploit the flexibility of this integrated energy system, advanced control strategies are needed.
This study proposes a comfort-oriented and grid-aware energy management system for an electrified building equipped with photovoltaics, a battery energy storage system, a hydrogen system, a heat pump, and domestic hot water storage.
A novel fuzzy logic-based controller coordinates fuel cell and heat pump operation while exploiting building thermal mass as a flexibility resource.
The proposed approach bridges the gap between simple rule-based control and model predictive control, enabling adaptive operation without system models, weather and demand forecasts, or high computational effort.
The controller is evaluated over twelve winter weeks for a German single-family household and benchmarked against a rule-based controller and an ideal-forecast model predictive controller.
Results show improved thermal comfort and lower electricity costs compared to the rule-based controller.
Compared to model predictive control, the proposed controller reduces fuel cell operating hours and toggling counts, leading to more practical system operation.
Overall, the proposed controller achieves a favorable trade-off between control performance and implementation complexity, combining low information requirements with improved comfort, storage use, and coordinated fuel cell-heat pump operation.

\end{abstract}


\end{frontmatter}


\setlength{\nomitemsep}{-\parskip} 
\makenomenclature
\renewcommand*\nompreamble{\begin{multicols}{2}}
\renewcommand*\nompostamble{\end{multicols}}

\ExplSyntaxOn
\NewExpandableDocumentCommand{\strcase}{mm}
 {
  \str_case:nn { #1 } { #2 }
 }
\ExplSyntaxOff

\renewcommand\nomgroup[1]{%
  \item[\bfseries
    \strcase{#1}{
      {P}{Parameters}
      {A}{Acronyms}
      {V}{Variables}
      {S}{Subscripts}
    }%
  ]%
}

\begin{table*}[h]
  \begin{framed}
    \nomenclature[A]{\acs{PV}}{\acl{PV}}
    \nomenclature[A]{\acs{EMS}}{\acl{EMS}}
    \nomenclature[A]{\acs{FLC}}{\acl{FLC}}
    \nomenclature[A]{\acs{MPC}}{\acl{MPC}}
    \nomenclature[A]{\acs{RBC}}{\acl{RBC}}
    \nomenclature[A]{\acs{RES}}{\acl{RES}}
    \nomenclature[A]{\acs{BESS}}{\acl{BESS}}
    \nomenclature[A]{\acs{COP}}{\acl{COP}}
    \nomenclature[A]{\acs{LOH}}{\acl{LOH}}
    \nomenclature[A]{\acs{SOC}}{\acl{SOC}}
    \nomenclature[A]{\acs{ECDF}}{\acl{ECDF}}
    \nomenclature[A]{\acs{CEMS}}{\acl{CEMS}}

%
    \nomenclature[V]{$P$}{Power in \si{\watt}}
    \nomenclature[V]{$T$}{Temperature in \si{\celsius}}
    \nomenclature[V]{$\dot{q}$}{Radiation in \si{\watt \per \square \metre}}
    \nomenclature[V]{$\dot{Q}$}{Heat flow in \si{\watt}}
    \nomenclature[V]{$\chi$}{Thermal state factor}
    \nomenclature[V]{$s$}{Binary variable}
    \nomenclature[V]{$p$}{Price in \si{\EUR/kWh}}
    \nomenclature[V]{$E$}{Energy in \si{k\watt\hour}}
    \nomenclature[V]{$\Delta$LOH}{Daily amount of hydrogen in \%}
    \nomenclature[V]{$\alpha$}{CEMS control output}
    \nomenclature[V]{WH}{Working hours}
    \nomenclature[V]{TOG}{Toggling counts}

%
    \nomenclature[P]{$\sigma$}{Costs in \si{\EUR}, \si{\EUR/\hour} or \si{\EUR/\watt\hour} }
    \nomenclature[P]{$t_\mathrm{s}$}{Sample time in \si{\hour}}
    \nomenclature[P]{$\eta$}{Efficiency in \%}
    \nomenclature[P]{$H$}{Number of time steps in the horizon}
    \nomenclature[P]{$N$}{Total number of time steps}
    \nomenclature[P]{$M$}{Number of time steps in a week}
    \nomenclature[P]{$k$}{Time step}

    \nomenclature[S]{$\mathrm{air}$}{Indoor air}
    \nomenclature[S]{$\mathrm{amb}$}{Ambient air}
    \nomenclature[S]{$\mathrm{w,in}$}{Internal wall}
    \nomenclature[S]{$\mathrm{w,out}$}{External wall}
    \nomenclature[S]{$\mathrm{we}$}{Week}
    \nomenclature[S]{$\mathrm{ch}$}{Charging}
    \nomenclature[S]{$\mathrm{d}$}{Discharging}
    \nomenclature[S]{$\mathrm{in}$}{Investment}
    \nomenclature[S]{$\mathrm{om}$}{Operation and maintenance}
    \nomenclature[S]{$\mathrm{buy}$}{Purchase}
    \nomenclature[S]{$\mathrm{sell}$}{Reimbursement}

%
%
    \printnomenclature
  \end{framed}
\end{table*}

\section{Introduction}
\label{sec:introduction}

Grid-aware buildings are increasingly recognized as an important component of future smart energy systems~\cite{tepe2025optimizing}. 
By adapting their energy consumption and on-site resources according to grid conditions, buildings can provide operational flexibility while reducing the need for energy imports \cite{woo2023comparing}.
To achieve decarbonization targets, grid-aware buildings are commonly equipped with distributed \acp{RES} such as \acf{PV} and wind turbines.
However, the variable nature of \ac{RES} generation introduces challenges in balancing local electricity production and energy demand \cite{cau2014energy}.
While \acf{BESS} can effectively address daily fluctuations between power generation and consumption, it is less suitable for mitigating seasonal mismatches because of its low energy density and high self-discharge rate \cite{zhang_comparative_2017}. 
To overcome this seasonal mismatch, hydrogen storage systems can be employed as a long-term storage solution, as they exhibit a low leakage rate and high energy density, storing energy in the form of hydrogen \cite{kaabinejadian2025systematic}.
Hydrogen system solutions in residential buildings are not considered optimal because of their high investment costs and low efficiencies. 
Although they are not yet economically viable, hydrogen systems have potential in resilient energy systems, enabling building energy autonomy and operation during power grid outages \cite{mena2024collective}. 
Moreover, as buildings have a significant impact on the transition to sustainability, integration of hydrogen systems combined with \ac{RES} can reduce greenhouse gas emissions, as addressed in \cite{naumann2024environmental}.
Beyond the environmental benefits, capturing waste heat from the fuel cell improves system efficiency by reducing the demand for other heating devices. 
In cogeneration mode, fuel cells can achieve overall efficiencies of 90\% as reported in the literature \cite{ellamla_current_2015,nguyen_proton_2020}. 
Furthermore, technological advancements have been shown to extend the operational lifespan of hydrogen systems \cite{ellamla_current_2015}, improving their economic viability. 
In addition, it is anticipated that economies of scale will result in a reduction of the investment costs associated with hydrogen systems \cite{liang2024deep, dodds_hydrogen_2015}.

Beyond the integration of fuel cells, the flexibility of residential energy systems can be further enhanced through electrified heat supply. 
In this context, the building thermal mass can act as an additional thermal storage resource. 
Combined with \ac{BESS}, hydrogen storage, and domestic hot water thermal storage, it enables better alignment of renewable generation and residential demand, thereby improving the security of the building's power supply \cite{wang_mpc-based_2024}.

In residential buildings equipped with diverse distributed energy resources, the \ac{EMS} plays a critical role in exploiting system flexibility and determining grid interaction \cite{shyni2024hess}. 
In the context of hydrogen-based buildings, grid-aware control strategies remain limited \cite{kaabinejadian2025systematic}.
In particular, control strategies that enhance flexibility by coordinating multiple distributed energy resources while keeping implementation complexity low are lacking. 
To address this gap, the present study proposes a novel grid-aware comfort-oriented energy management system for residential buildings that operates without weather forecasts, demand forecasts, or detailed energy component models.
Furthermore, it coordinates fuel cell and heat pump operation while using the building thermal mass as a source of flexibility.

\subsection{Related work}

Energy storage systems play a crucial role in improving self-sufficiency and minimizing operational costs for both large-scale energy systems \cite{YANG2024100068} and residential building applications \cite{go2023battery}. 
Among the many existing energy storage system configurations, \ac{BESS} for short-term storage and hydrogen for long-term storage are the most common combination in systems with high shares of \acp{RES}~\cite{LE2023120817, GIOVANNIELLO2023121311}. 
Using an office building, Li et al. \cite{li2022towards} propose a hybrid energy system comprising a \ac{BESS}, fuel cell, and electrolyzer. 
The developed game-theoretic power management system extends the fuel cell's operational lifespan and reduces power output fluctuations.
Furthermore, Go et al. \cite{go2023battery} investigate the role of combined \ac{BESS} and hydrogen storage in maximizing the self-sufficiency of residential building energy supply. 
In this sense, combining a \ac{BESS} with hydrogen storage improves temporal matching between renewable production and residential demand. 

Furthermore, fuel cell system efficiency improves when it is considered in cogeneration mode.
In this context, Ou et al. \cite{ou2021development} investigate efficient operation of a fuel cell, \ac{BESS}, and thermal storage. 
They highlight that heat recovery from fuel cell operation can increase overall system efficiency by 20\%. 
Gandiglio et al. \cite{gandiglio2014design} investigate the heat recovery potential of a residential 1\,\si{\kilo\watt} fuel cell in conjunction with an underfloor heating system. 
They achieve fuel cell system efficiency above 75\% in cogeneration mode.
Oh et al. \cite{oh2012optimal} evaluate the feasibility of introducing a fuel cell energy system in a residential apartment using a thermo-economic approach. 
The proposed cogeneration system yields 20\% operational cost savings compared to conventional systems. 
Herein, heat generated during fuel cell operation can be recovered and reused to meet the building's space heating and domestic hot water demands, while improving fuel cell efficiency.

Buildings with diverse energy resources depend heavily on how their components are managed.
The most common and widely used \ac{EMS} for residential building energy coordination are \ac{MPC} \cite{wang_mpc-based_2024}, \ac{FLC} \cite{vivas_fuzzy_2022}, \ac{RBC} \cite{vsanic2022stand}, and \ac{RL} \cite{SIEVERS2025100521}. 
\ac{MPC} offers a powerful control strategy and is often used to obtain optimization-based system behavior. 
However, the model dependency and relatively high computational cost limit its use \cite{comp_MPC, MAYNE20142967}. 
As a trade-off between simplicity and optimality, one solution is \ac{FLC}, an advanced form of \ac{RBC}. 
Soft computing capabilities, complex system management, and uncertainty handling make \ac{FLC} a strong candidate for governing the building's \ac{EMS}. 
Boynuegri et al. \cite{boynuegri2023real} study the load shifting potential of a residential \ac{EMS} using \ac{FLC}. 
The target is to reduce the hydrogen consumption of the off-grid house, which is supplied by renewable sources. 
They conclude that the \ac{FLC} is capable of reducing the hydrogen consumption of the fuel cell by 7.03\% and increasing the annual fuel cell efficiency by 4.6\% compared to a conventional \ac{EMS}, where the fuel cell is operated directly after the insufficiency of \ac{RES} power generation. 
However, the flexibility of \ac{BESS} and the fuel cell's heat recovery potential remain uninvestigated. 
Considering a single-family house, Šanić et al. \cite{vsanic2022stand} study the performance of a stand-alone household that combines cooling, heating, and power generation. 
With their \ac{EMS}, they can cover 80\% of the thermal loads by using heat recovered from the fuel cell and heat pump while operating in cooling mode. 
However, they do not explore the building's thermal flexibility. 
Vivas et al. \cite{vivas_fuzzy_2022} develop an \ac{FLC} and apply it to a residential building with \ac{BESS} and a hydrogen system. 
The proposed \ac{FLC} reduces \ac{BESS} and fuel cell degradation by 60\% compared to \ac{MPC}.
However, they use demand profiles for thermal and electrical consumption and do not explore the building's thermal flexibility. 
Duman et al. \cite{duman2021home} emphasize that \ac{EMS} and thermal control are commonly studied separately. 
Yet integrating them can yield benefits and enhance a building's overall flexibility.

\subsection{Contribution of the present paper}

The fuel cell and the heat pump were initially considered as competing technologies \cite{dodds_hydrogen_2015,knosala2022role}. 
However, in recent years, they have gained popularity as complementary technologies.
Andrade et al.~\cite{andrade_integrating_2025} show that integrating a heat pump and a fuel cell reduces fuel cell size by at least 40\%, while also decreasing initial investment costs. 
In contrast, the heat pump can contribute to 20\% energy savings. 
Despite these benefits, integrating multiple energy technologies increases the complexity of building energy management.
In particular, advanced control strategies often rely on detailed system models and forecasting methods, which can limit their practical implementation. 
Consequently, recent studies have highlighted the need for simpler control approaches to enable wider commercial implementation of building energy flexibility \cite{de2024living,amato2023room,clauss2024demonstrating}. Furthermore, integrating \ac{EMS} and thermal control has been identified as essential for improving flexibility \cite{duman2021home}.
Addressing these challenges, the present paper investigates the following research question:
\textit{How can energy management enable grid-aware operation of buildings with distributed energy resources through the coordinated control of fuel cells and heat pumps while minimizing dependence on predictive information and detailed system models?}

To address this research question, the main contributions of the present paper are the following: 
\begin{itemize}
    \item Development of a novel \acf{CEMS} for buildings that coordinates a fuel cell and a heat pump operation while exploiting building thermal mass as a flexibility resource.
    
    \item Proposal of a novel grid-aware control strategy with low information and sensor requirements, facilitating practical deployment in building energy systems.
    
    \item Benchmarking of the proposed controller against both a predictive model-based upper performance bound (\ac{MPC}) and a simple \ac{RBC}.

    \item Evaluation of the operational performance of coupled fuel cell and heat pump in a residential energy system, with particular emphasis on hydrogen use, heat recovery, grid dependence, occupants' comfort, and a detailed load analysis of the coordinated system under realistic dynamic electricity price signals.

\end{itemize}

\subsection{Structure of the present paper}
The remainder of this paper is organized as follows. Section \ref{sec:system_description} describes the system overview. 
Section \ref{sec:control} describes the control algorithms, while Section \ref{sec:parameters} provides the data, parameters, simulation setup, and evaluation metrics. 
Section \ref{sec:results} presents the results and Section \ref{sec:discussion} discusses them and outlines the assumptions and the limitations of the present work. 
Finally, Section \ref{sec:conclusion} concludes the paper and discusses the future work.
In addition, the supplementary material provides a detailed description of the models, \ac{CEMS}, and parameters used in the study. 
\section{System description}\label{sec:system_description}

\begin{figure}[t]
    \centering
    \includegraphics[width=\textwidth]{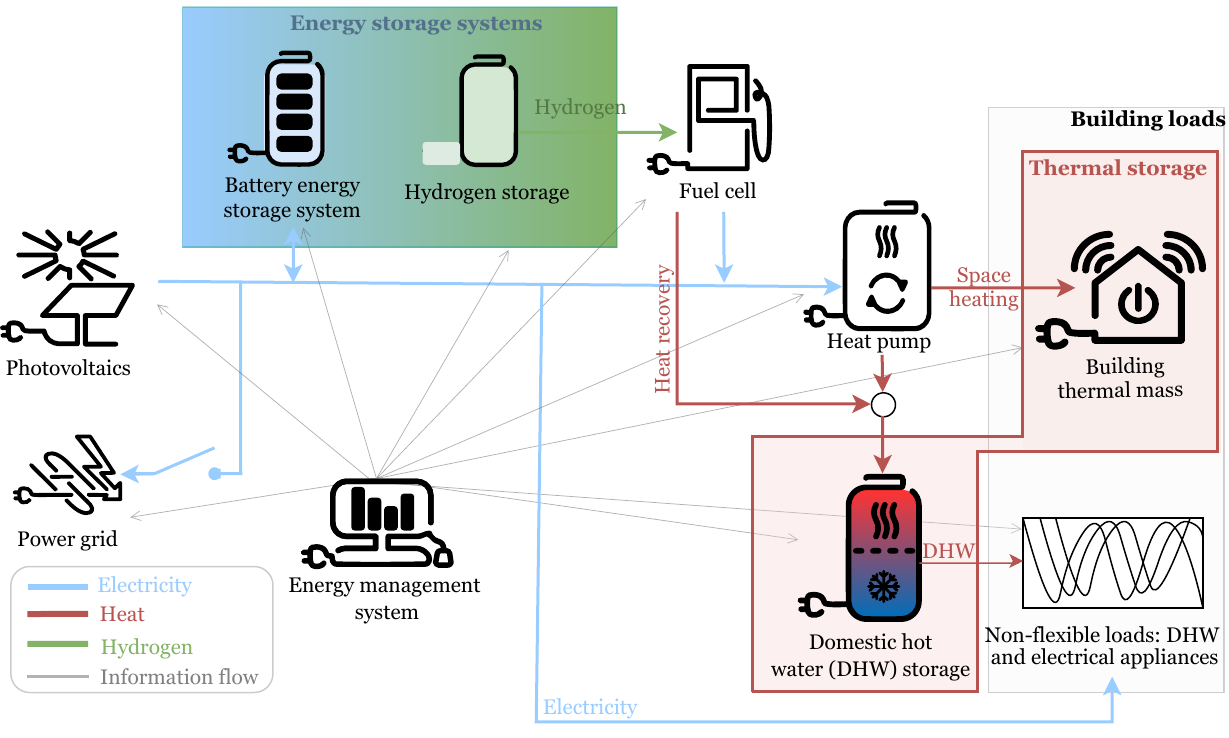}
    \caption{The residential building with \ac{PV}, \ac{BESS}, hydrogen storage, fuel cell, heat pump, domestic hot water storage, and building thermal mass storage.}
    \label{fig:setup}
\end{figure}

Figure \ref{fig:setup} illustrates the overview of the system, including all components and their corresponding electricity, heat, and hydrogen flows. 
The residential building represents a typical four-person household. 
From an energy source perspective, the power grid and \ac{PV} are considered.
They supply space heating and domestic hot water through a heat pump, as well as household electrical appliance loads. 
The power grid is represented by dynamic electricity price signals.
On the demand side, space heating is considered as a flexible demand with a predefined indoor comfort temperature range, allowing the building thermal mass to serve as thermal storage. 
In addition, domestic hot water storage provides additional thermal flexibility by decoupling heat production from consumption. 
In contrast, household electrical appliance loads are considered inflexible. 
To provide electrical flexibility through energy storage systems across multiple time scales, the system incorporates a hybrid energy storage architecture comprising a short-term \ac{BESS} and a long-term hydrogen storage.
The \ac{SOC} defines the state of stored energy in the \ac{BESS}, while the fill level of the hydrogen in storage is defined by the \ac{LOH}. 
For hydrogen-to-electricity conversion, this study selects the \ac{PEMFC} because it suits residential cogeneration, with high efficiency, fast start-up capability, and low-temperature operation \cite{byun2023reliability}. 
The produced heat can be used directly to meet the building's domestic hot water demand.
Hereafter, \ac{PEMFC} is referred to as the ``fuel cell”. 
It is assumed that the hydrogen storage is replenished with green hydrogen during the summer period using an electrolyzer.
This work focuses on the efficient use and management of stored hydrogen during winter periods with high energy demand.
Consequently, hydrogen production and electrolyzer operation are outside the scope of this study and are therefore not considered in the system modeling.

\subsection{Models}
\label{sec:model}

For the sake of conciseness, the detailed system modeling is provided in Section~S1 of the supplementary material.~\footnote{Note that all equations, sections, figures, and tables starting with ``S" are provided in the supplementary material.} 
It includes the mathematical formulations and parameter descriptions of all components, including the \ac{PV}, \ac{BESS}, hydrogen system, heat pump, domestic hot water storage, and thermal building model, along with the corresponding model assumptions. 

\subsection{Power balance}
Based on the considered energy components, the operational feasibility of the integrated energy system is ensured through an electrical power balance constraint, which enforces equality between the total generated and consumed power at each discrete time step $k$, expressed as:

\begin{equation}
    P_{\mathrm{PV},k} + P_{\mathrm{buy},k} + P_{\mathrm{FC},k} +  P_{\mathrm{d},k} = P_{\mathrm{sell},k} + P_{\mathrm{load},k}+ P_{\mathrm{HP},k} + P_{\mathrm{ch},k} \label{eq:power_balance}
\end{equation}

where $P_{\mathrm{PV},k}$ represents the \ac{PV} power generation, $P_{\mathrm{buy},k}$ purchased power from grid, $P_{\mathrm{FC},k}$ fuel cell power, $P_{\mathrm{d},k}$ \ac{BESS} discharging, $P_{\mathrm{sell},k}$ power sold to the grid, $P_{\mathrm{load},k}$ electrical load of the electrical appliances generated as a typical curve for a family household, $P_{\mathrm{HP},k}$ power consumed by the heat pump and $P_{\mathrm{ch},k}$ charging power of \ac{BESS}, where the power is measured in \si{\watt}.

\section{Control algorithms}
\label{sec:control}

This section describes the developed control algorithms. 
First, we describe the structure of the proposed algorithm \ac{CEMS}.
Next, we describe the lower benchmark \ac{RBC}, which requires minimal inputs, and the upper benchmark \ac{MPC}, which is an optimization-based control algorithm.
Finally, we compare and discuss the input requirements of all three controllers, as they mainly affect additional equipment and sensor installation.

\subsection{\acl{CEMS} (\acs{CEMS})} \label{sec:FLC}

The growing complexity of modern energy systems highlights the advantages of \ac{FLC}. Unlike model-based or forecast-dependent control strategies, \ac{FLC} is a soft-computing approach capable of effectively managing uncertainties, imprecision, and noisy inputs \cite{kaabinejadian2025systematic}.
Building parameters, such as indoor air temperature, or external parameters, such as electricity price signals, can be represented as soft linguistic variables that describe the current or future state of the system. 

\begin{figure}[h!]
    \centering
    \includegraphics[width=0.75\textwidth]{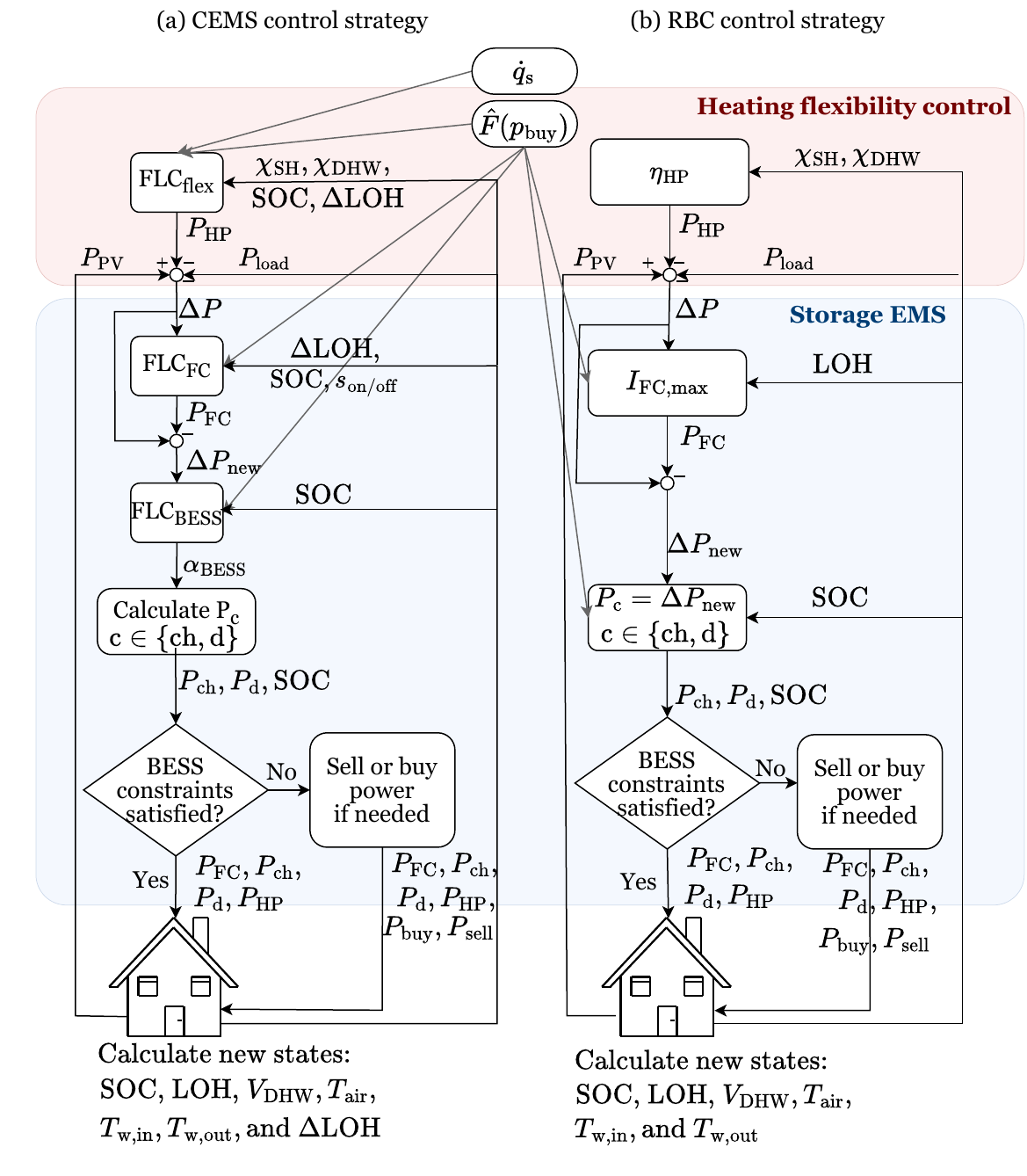}
    \caption{Information flow for: (a) \acf{CEMS} and (b) \acf{RBC}.}
    \label{fig:flc_rbc_flowchart}
\end{figure}

The idea behind the \ac{CEMS} is to design energy management without forecasts, using only predefined occupancy and electricity price signal profiles.
In this way, the controller does not depend on models or weather and demand forecasts and is simpler for practical implementation.
\ac{CEMS} hierarchically splits the flexible load management and operation of storage, as shown in Figure \ref{fig:flc_rbc_flowchart}(a). 
The flexible heating load is operated with FLC$_\mathrm{flex}$.
Its control actions are based on the following inputs: (i) SOC and $\Delta$LOH, (ii) current solar radiation $\dot{q}_\mathrm{s}$, (iii) indoor air temperature thermal state $\chi_\mathrm{SH}$ and domestic hot water storage thermal state $\chi_\mathrm{DHW}$, and (iv) \ac{ECDF} of the day-ahead electricity price signals, $\hat{F}(p_\mathrm{buy})$. 
Based on these inputs, FLC$_\mathrm{flex}$ determines the magnitude of the flexible heating load. 
The thermal states of space heating $\chi_\mathrm{SH}$ and domestic hot water $\chi_\mathrm{DHW}$ are defined as follows:
\begin{align}
    \chi_{\mathrm{SH},k} &= \sqrt{\frac{T_{\mathrm{air},k-1}-\text{max}(T_{\mathrm{air,min},k}, T_{\mathrm{air,min},k+5})}{T_{\mathrm{air,max},k}-\text{max}(T_{\mathrm{air,min},k}, T_{\mathrm{air,min},k+5})}}\label{eq:s_sh},\\
    \chi_{\mathrm{DHW},k} &= \left( \frac{V_{\mathrm{DHW},k-1}-V_\mathrm{DHW,min}}{V_\mathrm{DHW,max}-V_\mathrm{DHW,min}} \right)^2,\label{eq:s_dhw}
\end{align}

where $T_{\mathrm{air},k-1}$ is indoor air temperature from the last time step, $T_{\mathrm{air,min}}$ and $T_{\mathrm{air,max}}$ are the respective minimal and maximal indoor air temperature constraints defined by occupants.
In addition, FLC$_\mathrm{flex}$ uses the minimum temperature constraint over the next five hours, $T_{\mathrm{air,min},k+5}$, to preheat if necessary. 
Similarly, variables are defined for domestic hot water, where $V_{\mathrm{DHW},k-1}$ represents the volume of usable hot water in the storage from the previous time step, while $V_\mathrm{DHW,max}$ and $V_\mathrm{DHW,min}$ are maximal and minimal volume constraints, respectively.
The indoor air temperature state $\chi_\mathrm{SH}$ is under the square root to encourage temperature control near the minimum temperature constraint, reducing energy consumption. 
In contrast, the domestic hot water thermal state $\chi_\mathrm{DHW}$ is squared to keep the storage volume closer to its maximum and secure supply, since future demand is unknown. 
Furthermore, the \ac{ECDF} of electricity price signals is defined by the following equation:

\begin{align}\label{eq:EDF_prices}
    \hat{F}(p_{\mathrm{buy},k}) = \frac{1}{H}\sum_{i=k+1}^{k+H}I(X_i\le p_{\mathrm{buy},k}), 
\end{align}

where the condition $X_i\le p_{\mathrm{buy},k}$ represents the dynamic electricity prices that are less than or equal to the current electricity price signal $p_{\mathrm{buy},k}$, while $I(\cdot)$ indicates whether the condition is true.
The load defined by FLC$_\mathrm{flex}$ is first summed with the non-flexible electrical load, which comes from electrical appliances in the household. 
Then PV generation covers part or all of the load, if possible.  
If there is residual load, FLC$_\mathrm{FC}$ defines the amount that can be covered by the fuel cell based on: (i) residual load $\Delta P$, (ii) \ac{ECDF} of day-ahead electricity price signal $\hat{F}(p_{\mathrm{buy}})$, (iii) last fuel cell toggling state $s_{\mathrm{on/off},k-1}$, (iv) SOC and a percentage of allowable daily hydrogen consumption $\Delta$LOH.

The daily permissible amount of hydrogen $\Delta$LOH$_k$ depends on the previous state in the storage, the difference between the absolute hydrogen left in the storage $E_{\mathrm{H_2}}$ in two consecutive time steps and the initial daily allowable hydrogen use $\Delta E_{\mathrm{H_2,init},k}$ updated each new day:

\begin{align} 
\Delta \mathrm{LOH}_{k} &= 
\begin{cases} 
1, & \text{if } k \bmod H = 0, \\[1ex] 
\Delta\mathrm{LOH}_{k-1} + \dfrac{E_{\mathrm{H_2},k}-E_{\mathrm{H_2},k-1}} {\Delta E_{\mathrm{H_2,init}}}, & \text{otherwise}, \end{cases} 
\label{eq:deltaLOH_fuzzyEMS} \\
\Delta E_{\mathrm{H_2,init}} &= \dfrac{E_{\mathrm{H_2,}k}-E_\mathrm{H_2,min}}{\text{NoD}}  \qquad \qquad \quad \quad  \text{ for }  k \bmod H = 0.\label{eq:E_H2_init}
\end{align}

The variable $\Delta E_{\mathrm{H_2,init}}$ represents the initial amount of hydrogen that can be utilized on a given day. At the beginning of each day ($k \bmod H = 0$), Equation \eqref{eq:E_H2_init} determines this value by dividing the hydrogen energy available above the absolute minimum level, $(E_{\mathrm{H_2,}k}-E_{\mathrm{H_2,min}})$, equally across all remaining simulation days. 
In this way, the available hydrogen reserve is allocated uniformly throughout the simulation horizon, thereby preventing excessive consumption during the early days of operation. 
Furthermore, the daily hydrogen level $\Delta \mathrm{LOH}_k$ is initialized to 1 at the beginning of each day and updated according to the change in the absolute hydrogen content of the storage. Since this update is normalized by $\Delta E_{\mathrm{H_2,init}}$, $\Delta \mathrm{LOH}_k$ represents the remaining share of the daily hydrogen budget rather than an absolute quantity of hydrogen stored in the tank.
Parameter $E_{\mathrm{H_2,min}}$ denotes the minimum hydrogen energy that must remain in storage, while NoD represents the number of remaining simulation days.
In reality, the total number of days when hydrogen is required would be estimated based on when enough sunlight is available to use surplus \ac{PV} to produce hydrogen. 
The output of the FLC$_\mathrm{FC}$ is the power that the fuel cell can produce to meet demand. 

The \ac{BESS} is controlled with FLC$_\mathrm{BESS}$ using the following inputs: (i) residual of the required power to satisfy the demand fully $\Delta P_\mathrm{new}$ (ii) \ac{ECDF} of electricity price signal $\hat{F}(p_\mathrm{buy})$ and (iii) \ac{SOC}.
If \ac{BESS} cannot meet demand because of charging or discharging power constraints or because the storage is empty/full, the corresponding residual power is bought from or sold to the grid. 
The output of FLC$_\mathrm{BESS}$ is defined as factor $\alpha_\mathrm{BESS} \in [-1,1]$. 
Its sign indicates whether the \ac{BESS} is charging (positive) or discharging (negative). 
Section S2 in the supplementary material further describes the structure of FLC$_\mathrm{flex}$, FLC$_\mathrm{FC}$, and FLC$_\mathrm{BESS}$, including membership functions, rule bases, and details about control variables.

\subsection{Rule-based benchmark} \label{sec:RBC}

\ac{RBC} is based on the \ac{CEMS} hierarchical structure, and it requires a minimal number of inputs among compared controllers. 
Figure \ref{fig:flc_rbc_flowchart}(b) shows the information flow of \ac{RBC}. 
Namely, first the thermal states $\chi_\mathrm{SH}$ and $\chi_\mathrm{DHW}$ are defined. 
In this context, the same Equations \eqref{eq:s_sh} and \eqref{eq:s_dhw} are used as for the \ac{CEMS}, except that \ac{RBC} does not have inputs about the future temperature constraints, and both thermal states are compared equally. 
Furthermore, the indoor air temperature is controlled between 21 and 21.5\,\si{\celsius} to minimize energy consumption and avoid thermal discomfort.
The heat pump is operated in an on/off manner and is activated at 90\% of its rated power whenever heating demand is required.
After defining the flexible load and whether \ac{PV} generation can cover it, it is considered whether the fuel cell should be switched on. 
This decision depends on whether demand exists and whether the \ac{ECDF} of day-ahead electricity price signals is in the upper quartile (i.e., power is expensive at the current moment compared to day-ahead price signals). 
If so, the fuel cell turns on at maximum current $I_\mathrm{FC,max}$. 
The residual power is then calculated and used to charge or discharge the \ac{BESS}. 
However, if the \ac{ECDF} of day-ahead electricity price signals shows low prices, the \ac{BESS} is charged if needed and the load, if it exists, is purchased from the grid.

\subsection{Optimal benchmark}
\label{sec:MPC}

For the optimal benchmark, we choose an \ac{MPC} strategy that uses the component models (details in the supplementary material) to forecast building behavior. 
Based on this forecast, the \ac{MPC} solves an optimization problem to determine the control trajectory that minimizes the operation cost. 
The objective function comprises three parts: (i) the cost of power exchange with the grid, (ii) the investment and lifecycle costs of the \ac{BESS} and fuel cell, and (iii) the on/off costs of the fuel cell. 
In this case, the on/off costs represent fuel cell toggling, which also affects its lifespan. 
The objective function is defined as:

\begin{alignat}{3} &\min \quad \sum_{k=1}^{H} \Big[ \underbrace{ t_\mathrm{s} \left( p_{\mathrm{buy},k}P_{\mathrm{buy},k} - p_{\mathrm{sell}}P_{\mathrm{sell},k} \right) }_{\text{Grid exchange cost}} &{}+{}& \underbrace{ t_\mathrm{s} \left( \sigma_{\mathrm{BESS,ch}}P_{\mathrm{ch},k} + \sigma_{\mathrm{BESS,d}}P_{\mathrm{d},k} + \sigma_{\mathrm{FC}}s_{\mathrm{FC},k} \right) }_{\text{Lifecycle cost}} \notag\\ &&{}+{}& \underbrace{ \sigma_{\mathrm{FC,on}}s_{\mathrm{FC,on},k} + \sigma_{\mathrm{FC,off}}s_{\mathrm{FC,off},k} }_{\text{On/off cost}} \Big] \label{eq:mpc_objective} \\ &\text{s.t. Equations}\quad {\eqref{eq:power_balance}, \; (S1)\text{--}(S6),} \notag\\ &{(S11)\text{--}(S13), \; (S15)\text{--}(S25),} \notag \end{alignat}

where $p_\mathrm{buy}$ and $p_\mathrm{sell}$ represent the electricity and reimbursement price in \si{\EUR/k\watt \hour}, respectively, while the number of time steps in the prediction horizon is denoted as $H$. 
The $\sigma_\mathrm{B,ch}$, $\sigma_\mathrm{B,d}$ in \si{\EUR/k\watt\hour} and $\sigma_\mathrm{FC}$ in \si{\EUR/\hour} are the utilization costs consisting of the investment $\sigma_\mathrm{in}$ in \si{\EUR} and operation and maintenance costs $\sigma_\mathrm{om}$ in \si{\EUR/\hour}, calculated as following:  

\begin{align} \sigma_i &= \frac{1}{\eta_i} \left( \frac{\sigma_{\mathrm{in},i}}{L_i} + \sigma_{\mathrm{om},i} \right), \qquad i \in \{\mathrm{BESS,ch},\;\; \mathrm{BESS,d}, \;\;\mathrm{FC}\}, \label{eq:batt_andFC_operation_degradation} \\ L_{\mathrm{BESS},i} &= \frac{E_\mathrm{BESS,\max}}{P_i} N_{\mathrm{BESS}}, \qquad i \in \{\mathrm{ch}, \;\; \mathrm{d}\}, \label{eq:batt_lifecycle} \end{align}

where suffixes $\mathrm{BESS,ch}$ and $\mathrm{BESS,d}$ represent battery charging and discharging, respectively, while FC indicates the fuel cell.
The lifespans of the \ac{BESS} and fuel cell in hours are presented with $L_\mathrm{BESS}$ and $L_\mathrm{FC}$, respectively. 
Since the lifespan of the battery is not given by the datasheet, it is estimated based on depth-of-discharge cycles and the battery's capacity $E_{\mathrm{BESS, max}}$, as described in \cite{cau2014energy}.
Furthermore, we neglect the operation and maintenance costs of the \ac{BESS}.
The $\sigma_\mathrm{FC,on}$ and $\sigma_\mathrm{FC,off}$ are costs of toggling the fuel cell in \si{\EUR}, where $s_\mathrm{FC,on}$ and $s_\mathrm{FC,off}$ indicate whether the fuel cell is starting or stopping, defined by the following logic:

\begin{equation} \label{eq:on_off_logic}
\begin{aligned}
    s_{\mathrm{FC,on},k} &= s_{\mathrm{FC},k} \land \lnot s_{\mathrm{FC},k-1}, \\
     s_{\mathrm{FC,off},k} &= s_{\mathrm{FC},k-1} \land \lnot s_{\mathrm{FC},k}.
\end{aligned}
\end{equation}

The MPC employs all models described in the supplementary material as control models, except for the nonlinear model of the fuel cell. 
This nonlinearity causes convergence issues in the optimization, which we avoid by linearizing the relation between current and power output (see Figure S1 in the supplementary material). 
Since the horizon time is 24\,\si{\hour}, \ac{MPC} gets its permissible amount of hydrogen that it can utilize throughout the day by dynamically defining the LOH$_\mathrm{min,day}$ for each new day:
\begin{align}
    \Delta \text{LOH} &=\frac{\text{LOH}_\mathrm{max}-\text{LOH}_\mathrm{min}}{\mathrm{NoD}},\label{eq:deltaLOH_fuzzyEMS_mpc}\\
      \text{ LOH}_{\mathrm{min,day},k} &= 
      \begin{cases} \min\left(\text{LOH}_\mathrm{min},\text{ LOH}_{\mathrm{max}} -\Delta \text{LOH} \cdot \left(1+\frac{k}{H}\right)\right), \ &\text{if}\; k\text{ mod } H=0,\\
      \min(\text{ LOH}_{\mathrm{min}},\text{ LOH}_{\mathrm{min,day,}k-1}), \ &\text{otherwise,} 
      \end{cases}\label{eq:LOHmin_MPC}
\end{align}

where NoD represents the number of remaining days of the simulation,  LOH$_\mathrm{min}$ and LOH$_\mathrm{max}$ are the absolute minimal and maximal levels of the hydrogen in the storage, $H$ is the number of time steps in the horizon, and $k$ is the time step. 
Moreover, the initial level of hydrogen in the tank at the beginning of the simulation is equal to the maximal level of the hydrogen, i.e., LOH$_\mathrm{init}$ = LOH$_\mathrm{max}$.

\subsection{Inputs required for each controller} \label{sec:information}

Table \ref{tab:level_of_information} gives an overview of the required inputs by controllers.
The lower input requirements simplify implementation by reducing the need for additional sensors, equipment, and highly accurate forecasting models.
Thus, the crosses in the table are shown in green, highlighting the controller's lower input requirements.
All three controllers require the current indoor air temperature $T_\mathrm{air}$, volume of usable hot water in storage $V_\mathrm{DHW}$, and state of \ac{BESS} and hydrogen storage. 
\ac{CEMS} and \ac{MPC} also require the fuel cell's previous state and the permitted daily hydrogen use $\Delta$LOH.
The forecasting models include \ac{PV}, \ac{BESS}, hydrogen storage, fuel cell, domestic hot water storage, heat pump, and thermal building envelope models. 
Moreover, forecasting profiles include weather, domestic hot water thermal load, and appliances' electrical load. 
MPC requires both forecasting models and profiles. Day-ahead temperature constraints, $T_\mathrm{air,min}$ and $T_\mathrm{air,max}$, are user-defined temperature constraints required by MPC. 
\ac{CEMS} requires only 5 hours ahead of indoor air temperature constraints, which is just enough to preheat if necessary. 
Day-ahead electricity prices, which each controller requires, are not considered to be predictions, because the complete price profile for the next day is available in advance.
Thus, the dynamic electricity price signal is treated as a known external input from the consumer's perspective.
From current weather conditions, \ac{CEMS} uses only the solar radiation $\dot{q}_\mathrm{s}$ for taking actions about flexible heating.

\begin{table}[h]
  \centering
  \caption{Input requirements: \textbullet\ denotes a required input and \textcolor{KITgreen}{\ding{55}} a non-required input. Lower input requirements are considered advantageous. Forecasting models and profiles refer to energy system component models and weather and demand forecasting profiles, respectively.}
  \newcolumntype{C}[1]{>{\centering\arraybackslash}p{#1}}
  \begin{tabular}{C{1.9cm}| C{2cm} C{2cm} C{2cm} C{1.9cm} C{1cm} C{0.8cm}}
    Input 
      & $T_\mathrm{air}$ 
      & $V_\mathrm{DHW}$ 
      & SOC 
      & LOH 
      & $\Delta$LOH 
      & $s_\mathrm{on/off}$   \\
    \hline
    RBC 
      & \textbullet 
      & \textbullet 
      & \textbullet 
      & \textbullet 
      & \textcolor{KITgreen}{\ding{55}}  
      & \textcolor{KITgreen}{\ding{55}}  \\  
    \ac{CEMS} 
      & \textbullet 
      & \textbullet 
      & \textbullet 
      & \textbullet 
      & \textbullet 
      & \textbullet \\
    MPC 
      & \textbullet 
      & \textbullet 
      & \textbullet 
      & \textbullet 
      & \textbullet 
      & \textbullet \\
    \hline
    \hline
    \multirow{2}{=}{\centering Input} 
      & Forecasting models 
      & Forecasting profiles 
      & Future temp. constraints 
      & Day-ahead el. price 
      & \multirow{2}{1cm}{\centering $T_\mathrm{amb}$}
      & \multirow{2}{0.8cm}{\centering $\dot{q}_\mathrm{s}$}    \\
    \hline
    RBC 
      & \textcolor{KITgreen}{\ding{55}}  
      & \textcolor{KITgreen}{\ding{55}}  
      & \textcolor{KITgreen}{\ding{55}}  
      & \textbullet 
      & \textcolor{KITgreen}{\ding{55}}  
      & \textcolor{KITgreen}{\ding{55}}  \\  
    \ac{CEMS} 
      & \textcolor{KITgreen}{\ding{55}}  
      & \textcolor{KITgreen}{\ding{55}}  
      & \textbullet 
      & \textbullet 
      & \textcolor{KITgreen}{\ding{55}}  
      & \textbullet \\
    MPC 
      & \textbullet 
      & \textbullet 
      & \textbullet 
      & \textbullet 
      & \textbullet 
      & \textbullet \\
  \end{tabular}

  \label{tab:level_of_information}
\end{table}

\section{Evaluation setup}\label{sec:parameters}

This section describes the data, profiles, parameters, and evaluation metrics used in this study. 
We evaluate the controllers over twelve winter weeks in Germany, using weather and electricity price data from December 2022 to February 2023 for a base scenario. All models and control algorithms are implemented in Python, where the MPC utilizes the Gurobi optimizer \cite{gurobi}. 
The simulation resolution is 1\,\si{\hour} for all controllers, while the MPC prediction horizon is $24\,$\si{\hour}.
At the beginning of the simulation, the hydrogen storage is assumed to be filled up to the maximum with green hydrogen produced during the summer months, while the SOC is set to 50\%.

\subsection{Input data}

The demand profiles used in the present study are domestic hot water demand and electricity demand. 
The domestic hot water demand is generated using the DHWCalc tool \cite{DHW_load_profiles}, which uses statistical methods to create profiles. 
The electricity load of the appliances in the building is generated using the LPG tool \cite{pflugradt2022loadprofilegenerator} by setting the parameters for a typical four-person household, consisting of two parents and two children. 
The hourly dynamic electricity prices $p_\mathrm{buy}$ for Germany are obtained from aWATTar \cite{aWATTar}. 
They consist of a constant base price and a variable price signal defined for the day-ahead market. 
In contrast, the reimbursement price is constant $p_\mathrm{sell}=0.08\,$\si{\EUR/k\watt\hour} and is retrieved from the German Federal Network Agency \cite{selling_price}. 
The meteorological data are retrieved from the German Weather Service \cite{DWD}. 
In particular, the following factors are taken into consideration: the ambient air temperature $T_\mathrm{amb}$, global solar radiation $\dot{q}_\mathrm{s}$, and diffuse solar radiation. 

The occupancy schedule, used in \ac{MPC} and \ac{CEMS}, is further derived from \cite{ueno2020method}, where the schedule differs according to the day of the week and occupancy type, categorized as follows: 
 
\begin{align} 
T_{\mathrm{air,max},k} &= 24\,\si{\celsius}, \,\,\quad \forall k, \\ 
T_{\mathrm{air,min},k} &= 
\begin{cases} 
21\,\si{\celsius}, & \text{if } k \in \text{home},\\ 18\,\si{\celsius}, & \text{if } k \in \text{sleeping, away}, 
\end{cases} \\
\text{for all } k &\in \{1,\ldots,N\},\notag\\
\text{home} &= 
\begin{cases} 
[6\,\si{AM},8\,\si{AM}) \cup [5\,\si{PM},11\,\si{PM}), & \text{if weekday},\\ [6\,\si{AM},11\,\si{PM}), & \text{if weekend}, 
\end{cases} \\ 
\text{sleeping, away} &= 
\begin{cases} 
[11\,\si{PM},6\,\si{AM}) \cup [8\,\si{AM},5\,\si{PM}), & \text{if weekday},\\ [11\,\si{PM},6\,\si{AM}), & \text{if weekend}. 
\end{cases} 
\end{align}

On the other hand, the \ac{RBC} employs a hysteresis-based temperature control strategy. By maintaining the indoor air temperature between 21 and 21.5 \si{\celsius}, it reduces energy consumption and avoids occupant discomfort.

\subsection{Thermal building parameters}

The building envelope parameters are identified based on the Living Lab Energy Campus building on Campus North of the Karlsruhe Institute of Technology \cite{wiegel2022smart}. 
The building consists of two floors with five rooms, a kitchen, and a bathroom, with a total area of 100\,\si{\meter}$^2$. 
The parameter set is for highly insulated buildings, where the parameter identification and validation are evaluated during the German winter months and are described in detail in \cite{langner_model_2024}. 

The domestic hot water demand for a four-person household is 200\,\si{liter/day} based on \cite{DHW_200l}. 
Accordingly, the storage size is selected as $V_\mathrm{DHW,max} = 300\,\si{liter} $ to cover the demand. 
Standing losses are $\dot Q_\mathrm{l,DHW}=35\,\si{\watt}$ as the storage is chosen to be the A+ efficiency level according to EU regulation 814/2013 \cite{standing_losses}. 
The temperature of the storage $T_\mathrm{DHW}$ is set to $45\si{\degree C}$. 
The storage is not employed for direct utilization. Instead, it functions as a heat storage. 
Using a heat exchanger, the stored heat is transferred to the fresh water supply. 
Thus, the storage temperature can be reduced, avoiding potential hygiene-related issues \cite{dengiz2019demand}. 
Moreover, the heat recovery from the fuel cell is also used appropriately, since the water in storage is technical water, not drinking water \cite{warm_wasser_dhw}.

The heat pump chosen for this use case is the air-source LG Therma V Monobloc type HM051M U43 \cite{lg_heatpump} with a heating capacity of 5\,\si{k\watt}. 
As already mentioned, the domestic hot water storage is kept at $45\,\si{\degree C}$. 
In addition, space heating is set to $55\,\si{\degree C}$ using an underfloor heating system. 
The \ac{COP} varies from 1.93 to 5.16 for 45\,\si{\degree C} and from 1.6 to 4.15 for 55\,\si{\degree C} for outside temperatures $T_\mathrm{amb}$ between -15\,\si{\degree C} and 20\,\si{\degree C}.
Further information about parameters can be found in the supplementary material in Section S3.

\subsection{Renewable energy system}

\ac{PV} and \ac{BESS} parameters are adapted from the experimental building equipment. 
Moreover, the fuel cell parameters are based on \cite{cheng2025multi,wang2025dynamic,nondy20244e}. 
Full autarky would notably increase system size, since worst-case peaks would need to be covered.  
From an investment perspective, it is more cost-effective to use the grid during high-demand peaks. 
Therefore, we chose a hydrogen storage size that could cover half of the daily electricity demand over the winter months. 
The system efficiency of the fuel cell is assumed to be $\eta_\mathrm{FC}=75\%$ based on the study \cite{gandiglio2014design}. 
According to Stropnik et al. \cite{stropnik_influence_2022}, major degradation mechanisms of hydrogen components relevant to supervisory \ac{EMS} include on–off cycling and power fluctuations, while steady-state operation of the fuel cell can extend its lifespan by a factor of 12–18 compared to dynamic operation. 
Therefore, in addition to comfort- and cost-related indicators, the evaluation includes metrics related to fuel cell degradation, namely the number of switching events and operating hours.
All parameters considered in the renewable energy system are provided in the supplementary material in Section S3.

\subsection{Evaluation metrics} \label{sec:evaluation_metrics}

The comfort- and cost-oriented metrics utilized are thermal discomfort, electricity costs, and reimbursements.
These metrics exhibit a trade-off relationship where improving one typically leads to a decline in the other. 
The weekly thermal discomfort $d_\mathrm{we}$ is measured in \si{Kh} and calculated by multiplying the intensity of the temperature constraint violation by the duration of the violation \cite{mork2022nonlinear}: 

\begin{align} 
d_{\mathrm{we}} &= t_{\mathrm{s}} \sum_{k=1}^{M} d_k, \label{eq:discomfort}\\ 
d_k &= 
\begin{cases} 
T_{\mathrm{air,min},k} - T_{\mathrm{air},k}, & \text{if } T_{\mathrm{air},k} < T_{\mathrm{air,min},k},\\ T_{\mathrm{air},k} - T_{\mathrm{air,max},k}, & \text{if } T_{\mathrm{air},k} > T_{\mathrm{air,max},k},\\ 0, & \text{otherwise}. 
\end{cases} 
\end{align}

The weekly energy costs $\sigma_\mathrm{we,buy}$ and reimbursements $\sigma_\mathrm{we,sell}$ are calculated as the sum of the purchased $P_\mathrm{buy}$ or sold $P_\mathrm{sell}$ electricity multiplied by the current electricity price $p_\mathrm{buy}$ or reimbursement price $p_\mathrm{sell}$, respectively: 

\begin{align} 
\sigma_{\mathrm{we},i} &= t_{\mathrm{s}} \sum_{k=1}^{M} p_{i,k}\, P_{i,k}, \qquad \forall i \in \{\mathrm{buy},\,\mathrm{sell}\}, \label{eq:costs} \end{align}

where $M$ represents the number of time steps in a week. 

Furthermore, due to the high degradation of the fuel cell, we evaluate the total working hours and the number of times the device is switched on and off for all three controllers. 
The working hours are obtained by multiplying the binary variable $s_\mathrm{FC}$, which indicates the fuel cell is working,  by the sampling time $t_\mathrm{s}$ over $N$ time steps:  
\begin{align}
     \text{WH} &= t_\mathrm{s}\cdot \sum_{k=1}^N  s_{\mathrm{FC},k}.\label{eq:working_hours}
\end{align}
The total number of fuel cell on/off toggling events over the horizon of $N$ time steps is obtained by summing the binary variables $s_{\mathrm{FC,on},k}$, where $s_{\mathrm{FC,on},k}=1$ indicates switching on the fuel cell at time step $k$:
\begin{align}
         \mathrm{TOG} &=  \sum_{k=1}^N  s_{\mathrm{FC,on},k}\label{eq:on_off_switching}.
\end{align}

\section{Results}
\label{sec:results}

This section presents results comparing \ac{CEMS}, \ac{MPC}, and \ac{RBC} control actions. 
First, we compare the controllers' behavior by showing temperature profiles and storage use. 
Second, we compare comfort, cost, and reimbursement indicators weekly.
Then, we evaluate and discuss hydrogen use throughout the winter season and weekly heat recovery.
Finally, to evaluate the \ac{CEMS} robustness to parameter variation, we conduct sensitivity analysis.

\subsection{Demonstration of the controller behavior}
\begin{figure}[h!]
    \centering
    \includegraphics[width=0.9\textwidth]{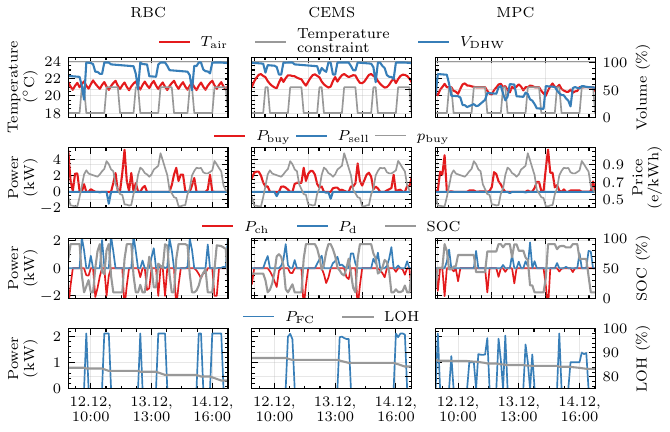}
    \caption{Example of a three-day performance comparison of all three controllers, taken from the twelve-week simulation. External input data are retrieved for the 2022-- 2023 years.}
    \label{fig:performance_comparison}
\end{figure}

Figure \ref{fig:performance_comparison} presents the exemplary three-day performance of all three controllers with high peaks of dynamic electricity prices. 
The upper plot row depicts the room air temperature and volume of the domestic hot water storage. 
Here, the gray lines represent the minimal and maximal air temperature constraints. 
The second row illustrates the power exchange with the grid and the electricity price signal. 
The last two rows show \ac{BESS} and hydrogen storage use, as well as the power consumed or provided to the corresponding system. 

\ac{MPC} keeps indoor air temperature mostly at the lower temperature constraint. 
At the same time, \ac{RBC}, as constructed, heats the room with hysteresis to minimize energy consumption, while \ac{CEMS} maintains a balanced temperature usually in the middle range. 
When it comes to the volume of the domestic hot water storage, the \ac{CEMS} and \ac{RBC} keep the volume more likely above 50\%  of the storage, since they do not forecast the domestic hot water demand.
In contrast, \ac{MPC}, having the perfect forecast, maintains the volume just enough to meet near-future demand. 
However, this situation is unrealistic because domestic hot water demand is inherently stochastic. 
Examining the plots of power exchange with the grid, all three controllers purchase from the grid when the electricity price signal is the lowest. 
However, \ac{MPC} and \ac{RBC} also show high peak purchases (see Dec. 13th for \ac{RBC} and Dec. 14th for MPC), which may also negatively impact the grid.

\ac{BESS} is utilized most frequently under \ac{RBC} and least frequently under \ac{MPC}. While the \ac{BESS} primarily operates as a daily energy storage system, hydrogen storage is employed less often, reflecting its role as a seasonal storage technology.

\subsection{Economic and comfort performance}

\begin{figure}[ht]
    \centering
    \includegraphics[width=0.9\textwidth]{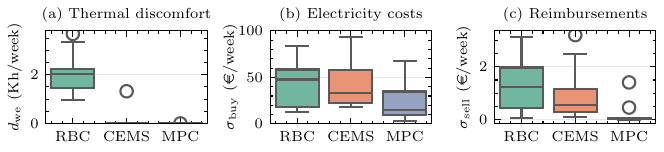}
    \caption{Weekly comfort violation and economic performance for the base scenario (year transition from 2022 to 2023).}
    \label{fig:quality_indicators_occupants}
\end{figure}

Economic and comfort-oriented indicators are evaluated weekly and presented in Figure~\ref{fig:quality_indicators_occupants} as box plots. 
Figure~\ref{fig:quality_indicators_occupants}(a) presents the weekly thermal discomfort for all three controllers. 
Among them, \ac{MPC} maintains occupant comfort throughout the entire evaluation period without any comfort violations. 
Similarly, \ac{CEMS} results in no thermal discomfort in 11 of the 12 analyzed weeks. In the remaining week, discomfort of 1.32\,~\si{\kelvin\hour} occurs because the controller prioritizes domestic hot water storage charging over space heating, as the domestic hot water profile is unknown. 
A similar effect occurs for \ac{RBC}, though more frequently, resulting in the higher median thermal discomfort of 2.01\,~\si{\kelvin\hour/week}. 
Nevertheless, this is considered low relative to the total duration of the week.

\ac{MPC} achieves the lowest median weekly electricity costs of 15.22\,\si{\EUR}, as depicted in Figure~\ref{fig:quality_indicators_occupants}(b). 
The \ac{CEMS} and \ac{RBC} have similar interquartile ranges for costs. 
However, \ac{CEMS} has a lower median value of 32.75\,\si{\EUR} compared to \ac{RBC} with 47.75\,\si{\EUR}. 
Figure~\ref{fig:quality_indicators_occupants}(c) shows that none of the three controllers receives notable reimbursements from selling energy. 
\ac{MPC} has the median of 0.04\,\si{\EUR/week}, \ac{CEMS} of 0.55\,\si{\EUR/week} and \ac{RBC} of 1.25\,\si{\EUR/week}. 
In general, \ac{MPC} attempts to use or store power generated from \ac{PV} for either thermal or electrical storage, as it does not find it optimal to sell it at the reimbursement price of $p_\mathrm{sell}=0.08\,$\si{\EUR/k\watt\hour}.
However, it shows two outliers in sold energy, corresponding to reimbursements of 0.46\,\si{\EUR/week} and 1.41\,\si{\EUR/week}, while \ac{CEMS} shows one outlier of 3.19\,\si{\EUR/week}. 
These outliers occur during periods of high solar radiation, when the system cannot store excess \ac{PV} generation and therefore sells it to the grid.
Because the reimbursement price is constant, sold energy is proportional to reimbursements.
Based on sold energy, \ac{MPC} has a median weekly sold energy of 0.51\,\si{k\watt\hour}, while \ac{CEMS} and \ac{RBC} have 6.96\,\si{k\watt\hour} and 15.69\,\si{k\watt\hour}, respectively. 
Comparing reimbursements with sold energy, \ac{RBC} gets only 0.7\,\si{\EUR} more weekly than \ac{CEMS} but sells 8.73\,\si{k\watt\hour} more energy than \ac{CEMS}, based on median values.

\subsection{Hydrogen storage use}

\begin{figure}[ht]
    \centering
    \includegraphics[width=0.9\textwidth]{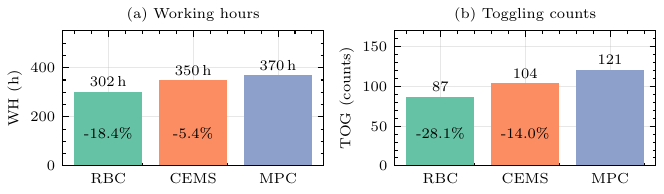}
    \caption{Total fuel cell working hours and toggling counts over a 12-week period for the base scenario (year transition from 2022 to 2023). Percentage is relative to the \ac{MPC} value.}
    \label{fig:working_hours_FC}
\end{figure}

Because the fuel cell serves as a seasonal energy storage system and its degradation is strongly influenced by operational behavior, Figure~\ref{fig:working_hours_FC} compares total working hours and toggling counts over the 12-week winter period. 
Interestingly, \ac{RBC} achieves an 18.4\% reduction in working hours and a 28.1\% reduction in toggling counts compared to the \ac{MPC}. 
\ac{RBC} sets the fuel cell to maximum power generation only when the electricity price signal is highest, while \ac{MPC} uses lower power generation based on energy demand. 
Accordingly, \ac{RBC} has fewer working hours and reduced toggling counts. 
\ac{CEMS} falls between \ac{MPC} and \ac{RBC} in fuel cell operation with reduction of 5.4\% for working hours and 14\% for toggling counts compared to \ac{MPC}. 

\begin{figure}[ht]
    \centering
    \includegraphics[width=0.49\textwidth]{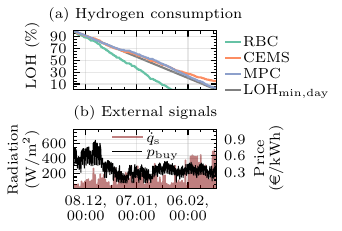}
    \caption{Hydrogen consumption in subplot (a) compared to external signals in subplot (b) throughout the entire evaluation period. External input data are retrieved for the 2022-- 2023 years.}
    \label{fig:hydrogen_timeseries}
\end{figure}

Figure~\ref{fig:hydrogen_timeseries}(a) depicts the hydrogen consumption over the entire period of the evaluation with external signals in Figure~\ref{fig:hydrogen_timeseries}(b). 
\ac{RBC} depletes the hydrogen storage fastest. 
Because information about the system is limited, \ac{RBC} uses all the hydrogen in the first two months of the simulation. 
\ac{MPC} considers the hydrogen consumption dynamically.
Around Jan. 7, \ac{MPC} uses less hydrogen than the permissible daily level because electricity prices are lower. 
\ac{CEMS} uses the hydrogen mostly linearly, with an impact of solar radiation.
It keeps more hydrogen supply in storage than the other two controllers at the end of the simulation because of a high \ac{PV} generation period in the last month. 
This relatively equal distribution of the hydrogen consumption comes from the fact that \ac{CEMS} considers only the relative electricity price signal $\hat{F}(p_\mathrm{buy})$. 
In that sense, \ac{CEMS} uses hydrogen when grid conditions are mostly unfavorable that day. 
In contrast, \ac{MPC} considers the absolute price signal, making decisions based on the current price $p_\mathrm{buy}$ and the day-ahead price distribution.
Furthermore, after a 12-week evaluation period, the \ac{LOH} left in storage by \ac{CEMS} and \ac{MPC} are 14.95\% and 3.97\%,  respectively. 
This can bring \ac{CEMS} closer to the \ac{MPC} by reducing electricity costs. 
However, it would also increase the fuel cell operation time.

\subsection{Heat recovery from fuel cell} \label{sec:heat_recovery}

\begin{figure}[h]
    \centering
    \includegraphics[width=0.9\textwidth]{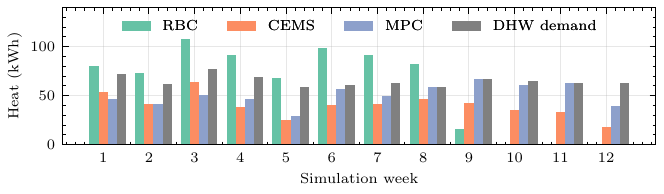}
    \caption{Weekly heat recovery from the fuel cell assuming a system efficiency of 75\% over the period from Dec. 1, 2022 to Feb. 22, 2023. \ac{RBC} has used up all its hydrogen reserves by week 9.}
    \label{fig:heat_recovery}
\end{figure}

\begin{figure}[h]
    \centering
    \includegraphics[width=0.45\textwidth]{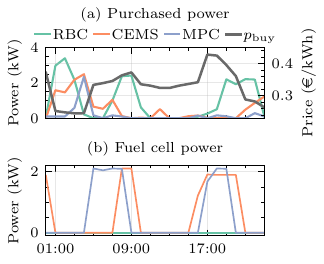}
    \caption{Power import from the grid (a) and fuel cell power (b) on Jan. 29, 2023.}
    \label{fig:power_exchange_with_grid_week9}
\end{figure}

In addition to electricity generation, the fuel cell also produces heat that can be recovered. 
Figure \ref{fig:heat_recovery} compares weekly heat recovery from the fuel cell for all three controllers. 
In addition, the figure shows weekly domestic hot water demand, including storage losses. 
In the first eight weeks, \ac{RBC} meets weekly demand only through fuel cell heat recovery. 
However, \ac{RBC} uses up all the hydrogen before the end of the evaluation period because it does not restrict the daily hydrogen use. 
\ac{CEMS} and \ac{MPC} do not cover the full weekly domestic hot water demand in most weeks. 
This is true except in weeks 8, 9 and 11, when \ac{MPC} can satisfy the entire domestic hot water demand with heat recovered from the fuel cell. 
For example, in week 9, the recovered heat is 67.11\,\si{k\watt\hour} and the domestic hot water demand is 66.85\,\si{k\watt\hour}. 
This week is characterized by an electricity price discrepancy of 0.1--0.54\,\si{\EUR/k\watt\hour} and a longer period of high price signals, during which \ac{MPC} avoids purchasing power from the grid. 
A similar effect occurs for \ac{CEMS}, which shifts grid purchases to the period with the lowest prices. Figure~\ref{fig:power_exchange_with_grid_week9} shows this behavior by comparing purchased power and fuel cell power between \ac{MPC} and \ac{CEMS}.

\subsection{Sensitivity analysis}

Based on the previous evaluation, \ac{CEMS} is an effective control strategy, achieving good performance with lower information requirements than \ac{MPC}.
To assess its suitability for real-world deployment, a sensitivity analysis is conducted to evaluate the controller's robustness under varying operating conditions. 
Table~\ref{tab:sensitivity_analyses} presents the considered scenarios: (i) the base scenario used for comparison with \ac{MPC} and \ac{RBC}; (ii) three additional dynamic electricity price signals from winter periods; (iii) constant temperature constraints ($T_\mathrm{air,min}=21\,\si{\celsius}$ and $T_\mathrm{air,max}=24\,\si{\celsius}$) instead of occupancy-based; (iv) a poorly insulated building in place of the highly insulated building model parameters \cite{langner_model_2024}; and (v) a combination of the poorly insulated building and constant temperature constraints.
The different scenarios are evaluated for weekly median thermal discomfort $d_\mathrm{we}$, weekly median electricity costs $\sigma_\mathrm{buy}$ and reimbursements $\sigma_\mathrm{sell}$, total fuel cell working hours WH and toggling counts TOG, and the hydrogen left in the storage LOH$_N$.

The defined temperature comfort range is violated only in the low-insulation scenario. 
However, a weekly median violation of 0.62\,\si{Kh} does not indicate significant discomfort.
Regarding electricity costs, reimbursements, and the remaining hydrogen in storage, the last two scenarios exhibit substantially larger deviations than the other cases. 
This is primarily due to poor thermal insulation, which increases heating demand, raises electricity costs significantly, reduces reimbursements compared to the base scenario, and fully depletes the hydrogen storage. 
The working hours and toggling counts remain close to the base scenario, with maximum deviations of 6.86\% and 14.42\%, respectively. 
Overall, the results indicate either small deviations from the base scenario or expected changes caused by poor building insulation and stricter temperature constraints. 
These findings indicate the robustness of the proposed \ac{CEMS} under the considered variations and constraints.

\begin{table}[h]
    \centering
    \includegraphics[width=0.8\textwidth]{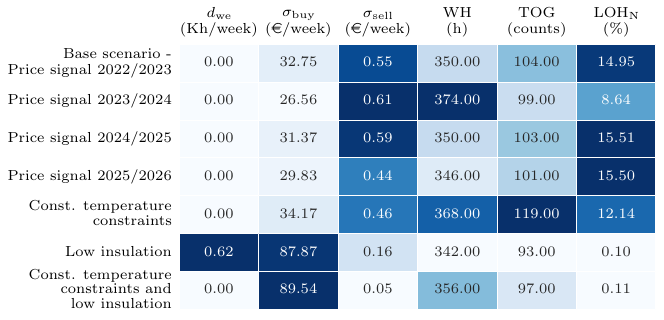}
    \caption{Impact of parameter variations on \ac{CEMS} performance.}
    \label{tab:sensitivity_analyses}
\end{table}

\section{Discussion}\label{sec:discussion}

The results show that \ac{CEMS} is more advantageous than \ac{RBC} in terms of thermal comfort, electricity costs, and reimbursements. 
\ac{CEMS} shows no temperature constraint violations in 11 of 12 weeks. 
It increases flexible thermal demand when the grid is in a favorable state or when \ac{PV} generation is high. 
This creates an opportunity to store flexible thermal energy in the building thermal mass, shifting the load from peak to off-peak hours. 
Considering exports to the grid, \ac{RBC} sells significantly more energy than \ac{CEMS} and \ac{MPC} because it operates the fuel cell at maximum power. 
After satisfying the demand, it charges the \ac{BESS} and sells the residual energy. 
However, \ac{MPC} does not find it optimal to sell the energy because the reimbursement price is too low, and instead retains it within the building energy system. 
For fuel cell operation, \ac{RBC} has the lowest working hours and toggling counts. 
This is advantageous because both affect fuel cell degradation. 
In this context, \ac{CEMS} performance falls between the two benchmarks with 350\,\si{h} of working hours and 104 toggling counts for three months of evaluation. 
For heat recovery, \ac{CEMS} and \ac{MPC} cannot meet weekly domestic hot water demand using only heat from the fuel cell in most of the weeks. 
In contrast, \ac{RBC} uses significantly more hydrogen at the start of the evaluation period and thus empties the hydrogen storage before the end of the evaluation period.
For \ac{CEMS}, solar radiation has the largest impact on daily hydrogen use, while for \ac{MPC}, the electricity price signal has a high impact.
\ac{CEMS} uses a relative price signal, built upon \ac{ECDF}, which lowers absolute price impact on the controller actions.
In contrast, \ac{MPC} considers an absolute day-ahead electricity price signal, which affects hydrogen use.

\subsection{Qualitative comparison of controllers}
Table \ref{tab:summary_performance} summarizes the qualitative comparison of the features and the performance of all three controllers.
Although \ac{MPC} can achieve the cost-optimal behavior, it is worth mentioning that it requires models and forecasting profiles, which makes \ac{MPC} a fragile solution, as some profiles are highly stochastic. 
Besides training and retraining the models, \ac{MPC} would require heat flow, current, voltage, and hydrogen flow sensors.
It also requires temperature sensors for all states in the building envelope model. 
In this case, besides the indoor air state, \ac{MPC} requires inner and outer wall temperature sensors (see Equations S23-S24 in the supplementary material). 
In this sense, \ac{CEMS} requires fewer inputs than \ac{MPC}, as it does not require forecasting models or profiles. 
However, it can offer a ``good-enough" solution rather than an optimal one. 
\ac{RBC} further reduces the required inputs compared to \ac{CEMS}, as it does not require inputs on predefined temperature constraints, solar radiation, the fuel cell's previous state, or the daily permissible amount of hydrogen to use.
The table compares discomfort, electricity costs, reimbursements, and heat recovery relative to the optimization-based \ac{MPC} behavior.
In contrast, fuel cell operation is evaluated using degradation indicators, where lower working hours and less toggling are preferred.

\begin{table}[t]
    \captionsetup{width=\linewidth}
    
    \begin{center}
    \caption{Summary of the controllers' performance and features. Poor performance is indicated by \poor, average performance by \avg, and good performance by \good. Discomfort, electricity costs, and reimbursements are relative to \ac{MPC}. Models- and forecast-free control refers to control without energy system component models as well as weather and demand forecasting profiles.}
    \label{tab:summary_performance}
    \renewcommand{\arraystretch}{1.1}
    \begin{tabular}{C{1.5cm} C{1.4cm} C{1.3cm} C{1.3cm}C{1.5cm} C{2.5cm} C{1cm}}   
       \multirow{2}{1.5cm}{Controller} & \multirow{2}{1.5cm}{\centering Model-free} & \multirow{2}{1.5cm}{\centering Forecast-free} &  \multirow{2}{1.3cm}{Number of inputs} & \multirow{2}{1.5cm}{\centering\ Hydrogen use}  & Thermal building mass use &\multirow{2}{1.5cm}{Exchange with grid}\\
       \hline
       
       \ac{RBC} & \good & \good &\good & \poor  & \poor & \poor\\
       
       \ac{CEMS} & \good & \good &\avg& \avg& \good& \avg\\
       
       \ac{MPC} & \poor & \poor &\poor & \good & \good & \good\\
       \hline
       \hline
       
       \multirow{3}{1.5cm}{Controller} &  \multirow{3}{1.5cm}{Discomfort} & \multirow{3}{1.5cm}{\centering Electricity costs} &  \multirow{3}{1.5cm}{\centering Reim- bursements} & Fuel cell working hours& \multirow{3}{1.3cm}{\centering Fuel cell toggling} & \multirow{3}{1cm}{\centering Heat recovery}\\
       \hline
       
       \ac{RBC} & \poor & \poor &\poor & \good  &\good & \poor \\
       
       \ac{CEMS} & \avg & \avg &\avg&\avg &\avg& \avg \\
       
       \ac{MPC} & \good & \good &\good & \poor & \poor&\good \\
    \end{tabular}
    \end{center} 
\end{table}

\subsection{Limitations}
\label{sec:limitations}

We do not consider minimum on/off times or ramp-rate constraints for system components. 
To enable fast model predictive control computation, we use simplified component models. 
Consequently, the model predictive control represents an idealized optimization benchmark based on near-perfect forecasts and near-optimal system operation. 
This benchmark is valuable for assessing the performance gap between the proposed controller behavior and an optimization-based behavior. 
It uses the same models for the evaluation and control, except for fuel cell operation, which is linearized. 
In addition, the analysis is limited to a single system configuration, including fixed photovoltaic parameters, storage capacities, and one geographic location. 
The underlying assumptions for each component model are summarized in Section~S1 of the Supplementary Material.

\section{Conclusions}
\label{sec:conclusion}

The increasing integration of distributed energy resources in residential buildings creates new opportunities for flexibility provision and peak management. 
While battery energy storage systems can effectively shift energy over short timescales, hydrogen storage offers additional potential for seasonal energy balancing. 
To realize these advantages, an energy management system is needed to coordinate multiple energy sources, storage systems, and flexible thermal loads while maintaining occupant comfort.

To address this challenge, this paper proposes a novel comfort energy management system for grid-aware coordination of battery and hydrogen energy storage systems, heat pump, fuel cell, thermal storage, photovoltaic generation, and building thermal mass.
The proposed algorithm does not use generation or demand forecasting profiles or energy-component models to make decisions.
The two main external signals considered by the controller are the day-ahead electricity price and the indoor comfort temperature range (e.g., temperature constraints) defined by the occupants.
To evaluate its performance, the proposed novel controller is compared with a simplified rule-based controller, serving as a lower benchmark, and with a model predictive controller representing an optimization-based upper benchmark based on perfect forecasts.
The proposed controller balances occupant comfort, energy costs, and efficient use of distributed energy resources while avoiding the high computational and forecasting requirements typically associated with optimization-based approaches. 
Overall, the proposed controller offers an attractive compromise between the simplicity of rule-based control and the performance of optimization-based control.

Future work will investigate the robustness of the proposed methodology across different system sizes, demand and generation profiles, and climatic conditions. 
In addition, fuel cell degradation costs are treated as fixed parameters. 
Further research should therefore examine the impact of degradation costs and identify economically optimal values under different electricity pricing profiles. 
The proposed controller is compared against one upper benchmark, a cost-minimizing model predictive controller. 
Additional comparisons with self-sufficiency-oriented or multi-objective model predictive controller formulations could provide further insights into controller performance. 
Finally, the study assumes that the hydrogen storage is replenished with green hydrogen during summer periods. 
Future investigations should evaluate hydrogen production and storage operation under increasing summer cooling demands, particularly in residential buildings in warmer climates affected by climate change.

\section*{Acknowledgments}
We acknowledge the financial support of the Helmholtz Association of German Research Centres (HGF) within the framework of the Program-Oriented Funding POF IV in the program Energy Systems Design (ESD, project numbers 37.12.01, 37.12.02 and 37.12.03). 

\section*{Declaration of generative AI and AI-assisted technologies in the writing process}
During the preparation of this work, the authors used ChatGPT and Microsoft Copilot in order to improve the presentation of the manuscript. After using these tools, the authors reviewed and edited the content and take full responsibility for the content of the publication.

\section*{Appendix}
This manuscript is complemented by supplementary material that describes the models, control algorithm, and parameters in detail.

\raggedright





\end{document}